\documentclass[12pt]{article}
\usepackage[left=2.5cm,top=2.50cm,right=2.5cm,bottom=2.50cm]{geometry}
\usepackage{mathrsfs}
\usepackage{amsmath,amssymb,latexsym,color,cancel,graphicx,bbm,colortbl}
\usepackage[english]{babel}
\usepackage{multirow}
\usepackage[latin1]{inputenc}
\usepackage{ragged2e}
\usepackage{cite}
\usepackage{graphicx}
\usepackage{placeins}
\usepackage{float}
\DeclareMathAlphabet{\mathpzc}{OT1}{pzc}{m}{it}
\begin{document}
\date{}

\title{SU(1,1) coherent states for the Dunkl- Klein-Gordon equation in its canonical form }
\author{M. Salazar-Ram\'irez$^{a}$\footnote{{\it E-mail address:} msalazarra@ipn.mx},  J.A. Mart\'inez-Nuño$^{a}$, MR Cordero-L\'opez $^{a}$ } \maketitle

\begin{minipage}{0.9\textwidth}
\small $^{a}$ Escuela Superior de C\'omputo, Instituto Polit\'ecnico Nacional,
Av. Juan de Dios B\'atiz esq. Av. Miguel Oth\'on de Mendiz\'abal, Col. Lindavista,
Alc. Gustavo A. Madero, C.P. 07738, Ciudad de M\'exico, Mexico.\\
\end{minipage}

\begin{abstract}
Using representation-theoretic techniques associated with the \( \mathfrak{su}(1,1) \) symmetry algebra, we construct Perelomov coherent states for the Dunkl--Klein--Gordon equation in its canonical form, which is free of first-order Dunkl derivatives. Our analysis is restricted to the even-parity sector and to the regime where the curvature constant \( R \) is much smaller than the system's kinetic energy. The equation under consideration emerges from a matrix-operator framework based on Dirac gamma matrices and a universal length scale that encodes the curvature of space via the Dunkl operator, thereby circumventing the need for spin connections in the Dirac equation.
\end{abstract}

%PACS:
%Keywords:
\section{Introduction}
The Dunkl operator, introduced by Charles Dunkl in 1989, is a differential-difference operator that combines the standard derivative with a reflection operator and a deformation parameter \( \alpha \). Defined as \( D_x^\alpha = \frac{d}{dx} + \frac{\alpha}{x}(1 - R) \), it acts differently on even and odd functions, making it a valuable tool for systems with discrete symmetries and parity effects \cite{Du1}. Originating from the non-standard quantization schemes proposed by Wigner and Yang, the Dunkl operator has become a central object in mathematical physics for problems where reflection symmetry plays a fundamental role \cite{Win,Yan,Gre}. It is widely applied in quantum mechanics, particularly in the analysis of deformed harmonic oscillators, Coulomb-like systems, and relativistic wave equations such as the Klein--Gordon and Dirac equations \cite{Sar,Ham}.

In recent years, the Dunkl formalism has been extended to more complex systems through generalizations such as the Jacobi--Dunkl operator and multi-parameter versions~\cite{Tri}. These developments have broadened its applicability to models involving non-constant curvature, noncommutative spaces, and quantum statistical mechanics~\cite{Has1,Ball,Has2,Mera,Mera1,Hoc,Hoc1}. Notably, the Dunkl operator often unveils hidden algebraic structures such as \( \mathfrak{su}(1,1) \), contributing to the exact or quasi-exact solvability of integrable models, including the Calogero--Moser--Sutherland system~\cite{Sal1,Sal2,Bri,Pol,Lap,Ques,Kha}. Its unique capacity to incorporate both parity and algebraic symmetry makes it a powerful framework for theoretical analysis and physical modeling.

The isotropic Dunkl oscillator in two dimensions is analyzed by constructing a Hamiltonian from two independent parabosonic oscillators involving Dunkl derivatives. The model is superintegrable and admits separation of variables in both Cartesian and polar coordinates. Exact solutions are expressed in terms of generalized Hermite, Laguerre, and Jacobi polynomials. The associated symmetry generators form the Schwinger--Dunkl algebra, an extension of \( u(2) \) that includes reflection operators. The overlap coefficients are given in terms of dual \( -1 \) Hahn polynomials~\cite{Gene}.

B. Hamil and B. C. Lütfüoglu investigate relativistic quantum systems within the framework of Dunkl operators, focusing on two prominent three-dimensional cases. They first solve the Klein--Gordon--Dunkl oscillator in both Cartesian and spherical coordinates, obtaining exact analytical solutions in terms of Laguerre and Jacobi polynomials. Subsequently, they analyze the Coulomb potential under the same formalism, deriving both the energy eigenvalues and the associated wave functions. Their results show that the Dunkl deformation induces modifications in the energy spectrum and alters the structure of the quantum states. Additionally, they explicitly derive fine-structure corrections resulting from this algebraic deformation~\cite{Ham}.

The Klein--Gordon equation is generalized using Dunkl derivatives, leading to exact solutions for eigenvalues and eigenfunctions in arbitrary dimensions. Applications include the \( d \)-dimensional harmonic oscillator and the Coulomb potential. In hyperspherical coordinates, the oscillator spectrum is expressed in terms of confluent hypergeometric functions. For the Coulomb case, both bound and scattering states are analytically solved, allowing for the computation of particle creation probabilities via the Bogoliubov transformation. The Dunkl deformation is shown to significantly modify both the spectral properties and structural characteristics of relativistic quantum systems~\cite{Ham3}.

P. Sedaghatnia \textit{et al.} investigate the quantum harmonic oscillator within the framework of the generalized Wigner-Dunkl formalism, employing creation and annihilation operators that incorporate reflection symmetries. They analyze the spectral properties of the Hamiltonian and derive the exact quantum propagator. Deformed coherent states are constructed, and their nonclassical features-including photon statistics, the Mandel parameter, and bunching or anti-bunching behavior-are systematically studied ~\cite{Seda2}.

In Ref.~\cite{Seda3}, the authors develop a matrix-operator algebra involving the Dunkl derivative and a universal curvature constant. They reformulate the Dirac equation without the use of spin connections and derive the Klein--Gordon equation in its canonical form, free of first-order Dunkl derivatives.

The organization of the paper is as follows. In Sec.~2, we present a brief mathematical overview of the Klein--Gordon equation modified by the Dunkl operator in a curved spacetime background, based on the formulation developed by Sedaghatnia \textit{et al.}~\cite{Seda3}. Special emphasis is placed on the canonical form of the equation, which eliminates first-order Dunkl differential operators. In Sec.~3, we focus on the Dunkl--Klein--Gordon equation with \( a(x) = e^{-R x^2} \) under even-parity conditions (\( \delta = 1 \)). Using an algebraic approach based on the irreducible representations of the Lie algebra \( \mathfrak{su}(1,1) \), we derive exact analytical expressions for the energy spectrum and corresponding eigenfunctions. We also construct the associated radial Perelomov coherent states of the \( SU(1,1) \) group and compute their time evolution, providing graphical representations of both. In Secs.~4 and 5, we extend the analysis to two additional even functions, \( a(x) = \frac{1 - R x^2}{1 + R x^2} \) and \( a(x) = \frac{\sinh(x\sqrt{R})}{x\sqrt{R}} \). Within the same algebraic framework, we obtain results analogous to those in Sec.~3. Final remarks and conclusions are given in Sec.~6.

\section{A Review of the Klein-Gordon equation in a curved space in the presence of Dunkl operator(KGDO)}
The covariant Dirac equation with Dunkl derivatives for a free spin-\(\tfrac{1}{2}\) particle (with \( \hbar = c = 1 \)) takes the following form in an \((n+1)\)-dimensional curved spacetime~\cite{Seda3}
\begin{equation}\label{DunkD1}
    i\gamma^\mu\left( D_\mu + \Gamma_\mu \right) \psi = m \psi,
\end{equation}
here, the Dirac matrices are denoted by \( \gamma^{\mu} \), with indices \( \mu = 0, 1, \ldots, n \), and the spin connections are represented by \( \Gamma_{\mu} \). It is understood that repeated indices are summed over. Squaring the Dirac equation~(\ref{DunkD1}) yields the following expression~\cite{Alha1}
\begin{equation}\label{SDD1}
\left[-\gamma^{\mu}\gamma^{\nu}D_{\mu}D_{\nu} - \gamma^{\mu}D_{\mu}\gamma^{\nu}D_{\nu} + \lambda\{\Omega,\gamma^{\nu}\}D_{\nu} + \lambda\gamma^{\mu}D_{\mu}\Omega - \lambda^{2}\Omega^{2}\right]\psi = m^{2}\psi,
\end{equation}
where \( \lambda \) is a dimensionless parameter and \( \Omega \) is a spacetime-dependent matrix. Thus, the curved-space generalization of the Klein--Gordon equation incorporating Dunkl derivatives can be written as
\begin{equation}\label{DDM1}
\left[ g^{\mu \nu} \left( D_\mu D_\nu +{^{D}\Gamma}_{\mu\nu}^{\sigma} D_{\sigma}\right) + m^2 \right]\psi = 0.
\end{equation}
An alternative algebraic framework can be constructed when the Klein--Gordon equation (\ref{SDD1}) is considered in its canonical form, i.e. without spinor coupling terms or first--order Dunkl derivatives
\begin{equation}\label{AMDI}
A\gamma^{\nu} - \lambda\{\Omega,\gamma^{\nu}\} = 0,\hspace{0.5cm} A\Omega = \frac{\lambda}{2}\{\Omega,\Omega\} + \lambda R\mathcal{I},
\end{equation}
in this framework, \( R \) can be interpreted as the constant scalar curvature of spacetime (not to be confused with the reflection operator \( R_\mu \)), while \( \mathcal{I} \) denotes a diagonal, spacetime-dependent matrix. As a consequence of the modified algebra~(\ref{AMDI}), Eq.~(\ref{SDD1}) takes the following transformed form
\begin{equation}\label{EDKC}
\bigg[g^{\mu\nu}D_{\mu}D_{\nu} - \lambda^{2}R\mathcal{I}\bigg]\psi = -m^{2}\psi,
\end{equation}
this expression represents the canonical form of the Klein--Gordon equation in curved spacetime incorporating the Dunkl operator, where both first-order Dunkl derivative terms and spinor component couplings are absent. The action of the Dunkl differential operator \( D \) in one-dimensional configuration space on a function \( f(x) \) with well-defined parity is given by
\begin{equation}\label{DUNKD1}
Df(x) = \frac{df(x)}{dx} + \frac{\alpha}{x}(1-R_{x})f(x),
\end{equation}
likewise, the following definition can be used
\begin{equation}\label{DUNKD2}
Df(x) = \frac{df(x)}{dx} + \frac{2\alpha}{x}\delta f(x),
\end{equation}
here, \( \delta = 1 \) if \( f(x) \) is an even function, and \( \delta = 0 \) if \( f(x) \) is odd. Following the results reported in Ref.~\cite{Seda3}, the Dunkl--Klein--Gordon equation~(\ref{EDKC}) in a \((1+1)\)-dimensional spacetime with a static metric can be written as
\begin{align}\label{DIFSOD}
\bigg\{
&\frac{d^2}{dx^2} + 2\left(\frac{\alpha}{x}-iE\frac{b(x)}{a(x)}\right)\frac{d}{dx} + \frac{E^2-m^2}{a^2(x)}-\frac{b^2(x)}{a^2(x)}E^2 - (4i\alpha\delta E)\frac{b(x)}{xa(x)} - \frac{a''(x)}{2a(x)} \\
&+\frac{1}{4}\left(\frac{a'(x)}{a(x)}\right)^2\nonumber +\frac{\alpha(\alpha+1)}{x^2}
\bigg\}\phi(x)=0,
\end{align}
here, the ansatz \( \psi(t,x) = e^{-iEt} \phi(x) \) has been applied to Eq.~(\ref{DIFSOD}), with \( \phi(x) \) possessing definite spatial parity \( \delta \). The Dunkl derivatives are given by \( D_\mu = (D_0, D_1) = \left( \frac{d}{dt}, \frac{d}{dx} + \frac{2\alpha}{x} \delta \right) \). To facilitate the analysis, a transformation of the form \( \phi(x) = W(x) \psi(x) \), where \( W(x) \) is chosen as an even function, is introduced. This transformation ensures that \( \phi(x) \) and \( \psi(x) \) share the same parity, with \( \delta = 1 \) corresponding to even functions and \( \delta = 0 \) to odd ones~\cite{Seda3}. The objective is to eliminate the first-derivative term in Eq.~(\ref{DIFSOD}), which leads to the following equation
\begin{equation}
\left[\frac{d^2}{dx^2} - \frac{a''(x)}{2a(x)} + \frac{1}{4} \left( \frac{a'(x)}{a(x)} \right)^2 + \frac{E^2 - m^2}{a^2(x)} + (4i\alpha E\sqrt{R})(1 - \delta)(x\sqrt{R})^{2\alpha - 1} + \frac{2\alpha}{x^2}\right]\chi(x) = 0,
\end{equation}
where the Dunkl parameter \( \alpha \) is assumed to be a half-odd integer, which ensures that \( W(x) \) is an even function, as required. Additionally, the identity \( b(x) = a(x)(x\sqrt{R})^{2\alpha} \) has been used. As a result, two well-defined parity solutions are obtained
\begin{equation}
\psi_{\pm}(t,x) = (|x|\sqrt{R})^{-\alpha}\exp\left( \frac{iE}{\sqrt{R}(2\alpha + 1)}(x\sqrt{R})^{2\alpha + 1} - iEt \right)\chi_{\pm}(x).
\end{equation}
When $\delta=1$, the even-parity wave function $\psi_+(t,x)$ is
\begin{equation}\label{EDSOE}
\left[ \frac{d^2}{dx^2} - \frac{a''(x)}{2a(x)} + \frac{1}{4} \left( \frac{a'(x)}{a(x)} \right)^2 + \frac{E^2 - m^2}{a^2(x)} + \frac{2\alpha}{x^2} \right] \chi_{+}(x) = 0,
\end{equation}
here, \( a(x) \) denotes an even function that satisfies the flat-space limit, such as \( a(x) = e^{-R x^2} \), \( a(x) = 1 + (R x^2)^{\alpha_2} \), \( a(x) = \frac{1 - R x^2}{1 + R x^2} \), \( a(x) = \cosh(x \sqrt{R}) \), \( a(x) = \frac{\sinh(x \sqrt{R})}{x \sqrt{R}} \), \( a(x) = \cos(x \sqrt{R}) \), or \( a(x) = \frac{\sin(x \sqrt{R})}{x \sqrt{R}} \), among others. Meanwhile, the solution \( \psi_-(t,x) \), corresponding to odd parity with \( \delta = 0 \), is given by
\begin{equation}
\left[\frac{d^2}{dx^2} - \frac{a''(x)}{2a(x)}+ \frac{1}{4} \left( \frac{a'(x)}{a(x)} \right)^2+ \frac{E^2 - m^2}{a^2(x)} + (4i\alpha E\sqrt{R})(x\sqrt{R})^{2\alpha - 1} + \frac{2\alpha}{x^2} \right]\chi_{-}(x) = 0.
\end{equation}

\section{SU(1,1) radial coherent states case $a(x)=e^{-R x^2}$}
In this section, we adopt an algebraic approach based on the irreducible representations of the Lie algebra \( \mathfrak{su}(1,1) \), which allows us to derive exact analytical expressions for the energy spectrum and the corresponding eigenfunctions. Furthermore, we construct the radial Perelomov coherent states associated with the \( SU(1,1) \) group for the specific case \( a(x) = e^{-R x^2} \). As a result, Eq.~(\ref{EDSOE}) takes the following form
\begin{equation}\label{DDKR1}
\left[\frac{d^2}{dx^2}+ \frac{2\mu}{x^2}- R^2 x^2+ (E^2 - m^2) e^{2Rx^2}+ R \ \right]\chi_{+}(x) = 0,
\end{equation}
for very small values of the curvature constant \( R \) (relative to \( E^2 - m^2 \)), Eq.~(\ref{DDKR1}) can be approximated as
\begin{equation}\label{DDKR2}
\left[\frac{d^2}{dx^2}+ \frac{2\mu}{x^2}+ \Lambda^2x^2+ (E^2 - m^2)\ \right]\chi_{+}(x) = 0,
\end{equation}
where \( \Lambda = \sqrt{2R(E^2 - m^2)} \). Under the variable transformation \( r = \Lambda x^2 \) and \( \chi_{+}(x) = r^{-1/4} F(r) \), Eq.~(\ref{DDKR2}) takes the form
\begin{equation}\label{DKSOR}
\left[-r^2\frac{d^2}{dr^2}-\frac{E^2 - m^2}{4\Lambda}r-\frac{1}{4}r^2 \right]F(r) = \left( \frac{\alpha}{2} + \frac{3}{16} \right)F(r).
\end{equation}
The \( \mathfrak{su}(1,1) \) generators can be explicitly constructed by applying the Schr\"odinger factorization method, starting from the ansatz
\begin{equation}\label{FACS}
\left[x\frac{d}{dx} + \mathscr{A}x + \mathscr{B}\right]\biggl[-x\frac{d}{dx} + \mathscr{C}x + \mathscr{F}\biggr]F(r) = \mathscr{G}F(r),
\end{equation}
through comparison of the expanded form of Eq.~(\ref{FACS}) with Eq.~(\ref{DKSOR}), the complex coefficients \( \mathscr{A} \), \( \mathscr{B} \), \( \mathscr{C} \), \( \mathscr{F} \), and \( \mathscr{G} \) can be determined as follows
\begin{align}
\mathscr{A}&=\pm\frac{1}{2}i  , \hspace{0.5cm}\mathscr{C}=\pm\frac{1}{2}i, \hspace{0.5cm} \mathscr{B}=\pm\frac{E^2-m^2}{4\Lambda}i-1, \hspace{0.5cm} \mathscr{F}=\pm\frac{E^2-m^2}{4\Lambda}i,\\
\mathscr{G}&=\left(\frac{\alpha}{2} + \frac{3}{16}\right)-\left(i\frac{E^2-m^2}{4\Lambda}-1\right)\left(i\frac{E^2-m^2}{4\Lambda}\right).
\end{align}
Therefore, the differential equation governing \( F(r) \) can be written in a factorized form as
\begin{equation}
\left[\mathscr{K}_{\pm}\pm1\right]\mathscr{K}_{\mp}=\mathscr{H}+\mathscr{L}\left[\mathscr{L}\pm i\right],
\end{equation}
here
\begin{equation}
\mathscr{H}=\left(\frac{\alpha}{2} + \frac{3}{16}\right), \hspace{0.5cm} \mathscr{L}=\left(\frac{E^2-m^2}{4\Lambda}\right),
\end{equation}
with complex Schr\"odinger operators expressed by
\begin{equation}
\mathscr{K}_{\pm}= \mp r \frac{d}{dr} + \frac{1}{2}ir +i\left(\frac{E^2-m^2}{4\Lambda}\right).
\end{equation}
These results allow us to introduce two novel complex operators, defined as
\begin{equation}\label{OPKPM0}
\mathscr{D}_{\pm}=\mp r \frac{d}{dr} + \frac{1}{2}ir +\mathscr{Z}_3,
\end{equation}
where $\mathscr{Z}_3$ is a complex operator constructed from Eq.~(\ref{DKSOR}) and is given by
\begin{equation}\label{OPZ3}
\mathscr{Z}_3=i\left[ r \frac{d^2}{d r^2} + \frac{\left(\frac{\alpha}{2} + \frac{3}{16}\right)}{r}  + \frac{1}{4} r \right]=-i\left(\frac{E^2-m^2}{4\Lambda}\right).
\end{equation}
Through straightforward calculation, one finds that Eqs.~(\ref{OPKPM0}) and (\ref{OPZ3}) define operators that close the \( \mathfrak{su}(1,1) \) algebra
\begin{equation}
[\mathscr{Z}_3,\mathscr{D}_{\pm}]=\mp\mathscr{D}_{\pm}, \hspace{0.5cm} [\mathscr{D}_{-}, \mathscr{K}_{-}]=2\mathscr{Z}_3.
\end{equation}
To determine the energy spectrum of the Klein-Gordon equation in curved spacetime incorporating the Dunkl operator, we employ the unitary irreducible representations of the \( \mathfrak{su}(1,1) \) Lie algebra~\cite{Bar1,Pere}
\begin{equation}\label{OPTLA}
\mathscr{Z}_3|k, n\rangle = (k + n) |k, n\rangle,
\end{equation}
the eigenvalue equation associated with the quadratic Casimir operator \( \mathbb{C}^2 \) takes the form
\begin{equation}\label{CASC1}
\mathbb{C}^2 = -\mathscr{K}_+ \mathscr{K}_- + \mathscr{Z}_3(\mathscr{Z}_3 - 1) =-\left(\frac{\alpha}{2}+\frac{3}{16}\right)= k(k - 1),
\end{equation}
with the Bargmann index given by the expression
\begin{equation}\label{BARG}
k=\frac{1}{2}+\frac{\sqrt{1-8 \alpha}}{4},
\end{equation}
This leads us to an important relationship derived from Eqs.(\ref{OPZ3}), (\ref{OPTLA}) and (\ref{BARG})
\begin{equation}
\frac{1}{2}+\frac{\sqrt{1-8 \alpha}}{4}+n=-i\left(\frac{E^2-m^2}{4\sqrt{2R(E^2 - m^2)}}\right),
\end{equation}
therefore, the resulting expression for the energy spectrum takes the form
\begin{equation}
E_{n}^{\,2} = m^{2} - 8R \left( 2n + 1 + i \sqrt{2\alpha - \frac{1}{4}} \right)^{2}.
\end{equation}
The energy spectrum \( E_n^2 \) is found to be complex, exhibiting a small negative imaginary part. This behavior indicates that the corresponding states are not bound in the conventional sense, but rather correspond to scattering or resonant states that decay gradually over time~\cite{Seda3}. The wave function \( F(r) \) associated with Eq.~(\ref{DKSOR}) can be derived from the corresponding differential equation\cite{Magh}
\begin{equation}
\frac{d^{2}F(r)_{nml}}{dr^{2}} + \left(\frac{\mathfrak{A}_{r}}{r} + \frac{\mathfrak{B}_{r}}{r^{2}} + \mathfrak{C}_{r}\right)F(r)= 0,
\end{equation}
resulting in the particular solution
\begin{equation}
F(r) = r^{\frac{1}{2} + \sqrt{\frac{1}{4} - B_{r}}} \, e^{-\sqrt{-C_{r}}\,r} \, L_{n}^{2\sqrt{\frac{1}{4} - B_{r}}}\left(2\sqrt{-C_{r}}\,r\right),
\end{equation}
where
\begin{equation}
\mathfrak{A}_{r} =\frac{E^2 - m^2}{4\Lambda{r}} ,\hspace{0.5cm} \mathfrak{B}_{r} =\left( \frac{\alpha}{2} + \frac{3}{16} \right), \hspace{0.5cm} \mathfrak{C}_{r} = \frac{1}{4}.
\end{equation}
Thus, the radial eigenfunction can be expressed explicitly as\cite{Gur}
\begin{equation}
F(x) = \sqrt{ \frac{2\, \Lambda^{\sigma + 1}\, n!}{\Gamma(n + 2\sigma + 1)} }\left( \sqrt{\Lambda} x \right)^{2\sigma + 1}\exp\left( -\tfrac{1}{2} i \Lambda x^2 \right)L_n^{2\sigma}\left( i \Lambda x^2 \right),
\end{equation}
with \( \sigma = \sqrt{\tfrac{1}{16} - \tfrac{\alpha}{2}} = k - \tfrac{1}{2} \). For the specific choice of \( a(x) \) considered in this section, we construct the Perelomov \( SU(1,1) \) coherent states as defined in Ref.~\cite{Pere}
\begin{equation}\label{defper}
|\varsigma\rangle=D\left(\xi\right)|k,0\rangle=\left(1-|\xi|^2\right)^k\sum^\infty_{n=0}\sqrt{\frac{\Gamma\left(n+2k\right)}{n!\Gamma\left(2k\right)}}\xi^n|k,n\rangle,
\end{equation}
therefore the radial coherent for the canonical Dunkl-Klein-Gordon equation in curved spacetime are
\begin{equation}
R_{n k}\left(x, \xi\right) =\left[\frac{2\Lambda^{k+\frac{1}{2}} \left(1 - |\xi|^2 \right)^{2k}}{ \Gamma\left(2k\right) } \right]^{1/2}\left( \sqrt{\Lambda}\, x \right)^{2k}\frac{ \exp\left[ \frac{i\Lambda x^2}{2} \left( \frac{\xi + 1}{\xi - 1} \right) \right] }{ (1 - \xi)^{2k} }.
\end{equation}
\begin{figure}[htbp]
    \centering
    \includegraphics[width=0.65\textwidth]{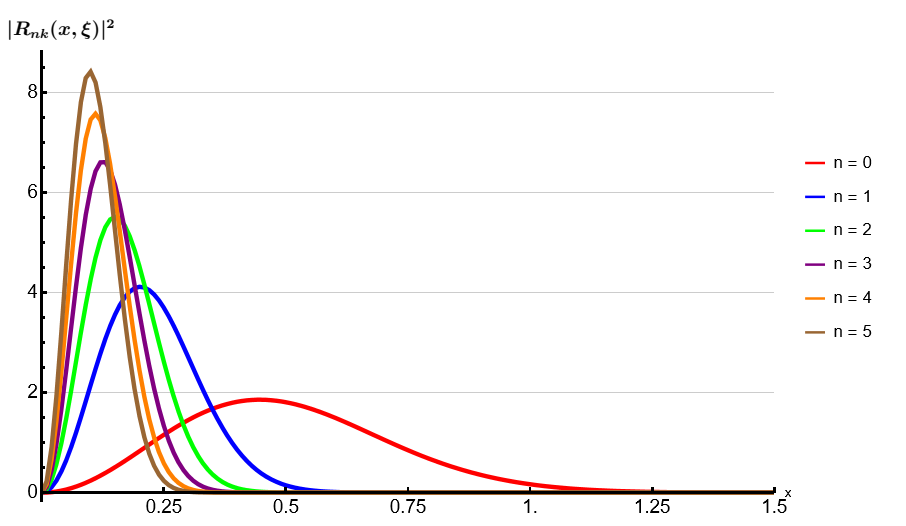}
    \caption{Normalized radial probability density $|R_{nk}(x, \xi)|^2$ associated with coherent states, for \( \alpha = \frac{1}{2} \), \( \xi = 0.5 + 0.2i \), and \( n = 0, 1, \dots, 5. \)}
\end{figure}

\begin{figure}[htbp]
    \centering
    \includegraphics[width=0.65\textwidth]{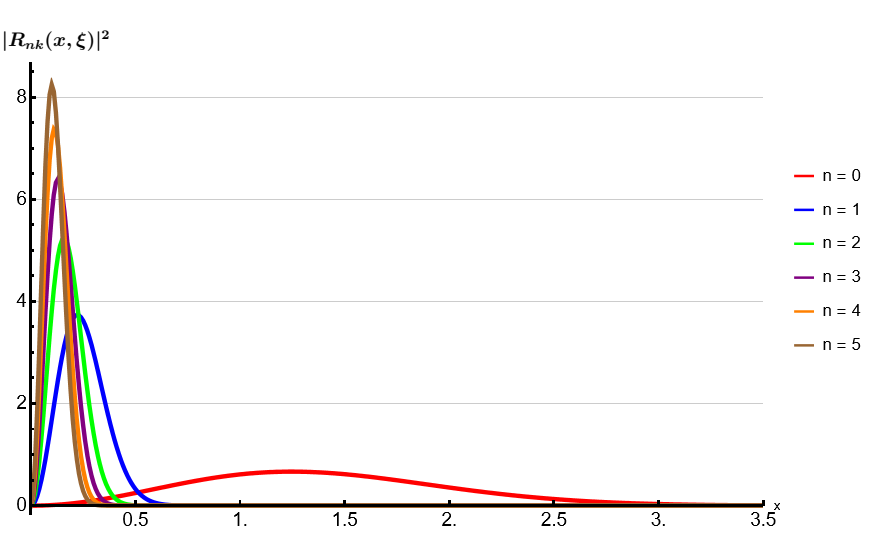}
    \caption{Normalized radial probability density  $|R_{nk}(x, \xi)|^2$  associated with coherent states, for \( \alpha = \frac{3}{2} \), \( \xi = 0.5 + 0.2i \), and \( n = 0, 1, \dots, 5. \)}
\end{figure}
\FloatBarrier
\begin{figure}[H]
    \centering
    \includegraphics[width=0.65\textwidth]{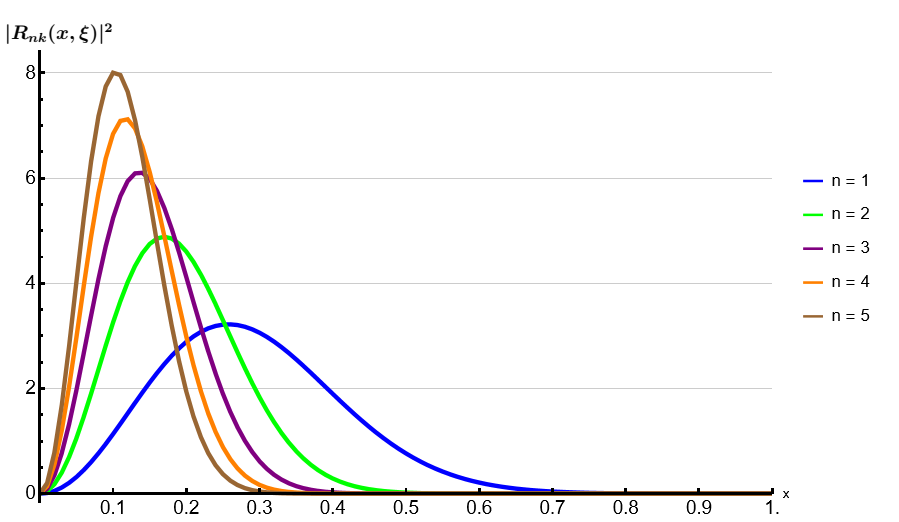}
    \caption{Normalized radial probability density  $|R_{nk}(x, \xi)|^2$  associated with coherent states, for \( \alpha = \frac{7}{2} \), \( \xi = 0.5 + 0.2i \), and \( n = 0, 1, \dots, 5. \)}
\end{figure}
The case \( n = 0 \) with \( \alpha = \frac{7}{2} \) is excluded from the analysis due to the numerically unstable behavior of the radial density \( |R^1_{nk}(x, \xi)|^2 \), which remains negligible over most of the physical domain and exhibits a sharp, nonphysical peak near the upper boundary. This behavior originates from the complex exponent \( k \) and the interference between exponential and power-law terms, resulting in a profile that lacks clear physical interpretation and dominates the plotted range.

Figures~1, 2, and 3 display the normalized radial probability densities \( |R_{nk}(x, \xi)|^2 \), corresponding to coherent states of the \( SU(1,1) \) group, for deformation parameters \( \alpha = \tfrac{1}{2}, \tfrac{3}{2}, \tfrac{7}{2} \), quantum numbers \( n = 0, \ldots, 5 \), and fixed complex parameter \( \xi = 0.5 + 0.2i \). These densities exhibit a sensitive dependence on both \( \alpha \) and \( n \). As \( \alpha \) decreases, the profiles become more concentrated near the origin, with sharper and more localized peaks. In contrast, increasing \( \alpha \) shifts the peak toward larger values of \( x \), producing slightly more extended distributions and indicating reduced radial confinement.

Additionally, for fixed \( \alpha \), increasing the quantum number \( n \) results in broader distributions and a gradual displacement of the maximum away from the origin, consistent with higher excitation and greater spatial extension. Altogether, these results demonstrate that the quantum localization properties of the coherent states are governed by a subtle interplay between the algebraic deformation encoded in \( \alpha \) and the excitation level determined by \( n \).

\subsection{Time evolution of the coherent states in the case $a(x)=e^{-R x^2}$}
We now focus on the computation of the time evolution of coherent states associated with the canonical  Dunkl-Klein-Gordon equation in curved spacetime. To this end, Eq.~(\ref{DKSOR}) can be written as follows
\begin{equation}\label{HET1}
\mathcal{H}_rF(r)=-i\left(\frac{E^2-m^2}{4\Lambda}\right)F(r),
\end{equation}
here
\begin{equation}\label{second4}
\mathcal{H}_r=i\left[ r \frac{d^2}{d r^2} + \frac{\left(\frac{\alpha}{2} + \frac{3}{16}\right)}{r}  + \frac{1}{4} r \right],
\end{equation}
by combining Eqs.~(\ref{OPZ3}) and (\ref{HET1}), one obtains the following result
\begin{equation}\label{OPB3ET}
\mathcal{H}_rF(r)=\mathscr{Z}_3F(r)=-i\left(\frac{E^2-m^2}{4\Lambda}\right)F(r).
\end{equation}
From this final expression, one can recover the energy spectrum previously obtained for the specific choice of $a(x)=e^{-R x^2}$
\begin{equation}
E_{n}^{\,2} = m^{2} - 8R \left( 2n + 1 + i \sqrt{2\alpha - \frac{1}{4}} \right)^{2},
\end{equation}
where the action of the operator $\mathscr{Z}_3$ on the $SU(1,1)$ states has been used explicitly~\cite{Sal1}. We define the time evolution operator associated with a general Hamiltonian as~\cite{Coh}
\begin{equation}
\Omega(\tau)=e^{-i\mathcal{H}_{r}\tau/\hbar}=e^{-i\mathscr{Z}_3\tau/\hbar},
\end{equation}
for a general Hamiltonian, the time evolution operator is given by~\cite{Gur,Gerry,NOS1}
\begin{equation}\label{PERET}
|\zeta(\tau)\rangle =\Omega(\tau)|\zeta\rangle=\Omega(\tau)\mathcal{D}(\xi)\Omega^\dag(\tau)\Omega(\tau)|k,0\rangle,
\end{equation}
where \( \tau \) denotes a fictitious time and \( \mathcal{D}(\xi) = \exp(\xi \mathscr{K}_{+} - \xi^{*} \mathscr{K}_{-}) \) is the displacement operator, with \( \xi \in \mathbb{C} \). Consequently, the time evolution of the lowest normalized state \( |k,0\rangle \), determined using Eq.~(\ref{OPTLA}), is
\begin{equation}\label{evest1}
\Omega(\tau)|k,0\rangle=e^{-ik\tau/\hbar}|k,0\rangle.
\end{equation}
The time evolution of the ladder operators is derived using the Baker-Campbell-Hausdorff (BCH) formula, as outlined below
\begin{equation}
\mathscr{K}_{+}(\tau)=\Omega^\dag(\tau)\mathscr{K}_{+}\Omega(\tau)=\mathscr{K}_{+}^{i\tau/\hbar},\hspace{0.5cm} \mathscr{K}_{-}(\tau)=\Omega^\dag(\tau)\mathscr{K}_{-}\Omega(\tau)=\mathscr{K}_{-}e^{-i\tau/\hbar}.
\end{equation}
Accordingly, the similarity transformation governing the operator dynamics in Eq.~(\ref{PERET}) is given by
\begin{equation}\label{opd1}
\Omega(\tau)\mathcal{D}(\xi)\Omega^\dag(\tau)=e^{\xi \mathscr{K}_{+}(-\tau)-\xi^*\mathscr{K}_{-}(-\tau)}=e^{\xi(-\tau)\mathscr{K}_{+} - \xi(-\tau)^*\mathscr{K}_{-}},
\end{equation}
where \( \xi(\tau) = \xi\, e^{\frac{i\tau}{\hbar}} \). Following the same procedure, when \( \zeta(\tau) = \zeta\, e^{\frac{i\tau}{\hbar}} \), the time-evolved displacement operator can be written in its normal-ordered form as
\begin{equation}\label{opedes}
\mathcal{D}(\xi(\tau))=e^{\zeta(-\tau) \mathscr{K}_{+}}e^{\eta \mathscr{Z}_3}e^{-\zeta(-\tau)^*\mathscr{K}_{-}}.
\end{equation}
Combining Eqs.~(\ref{evest1}) and~(\ref{opedes}), the resulting time-dependent Perelomov coherent state reads
\begin{equation}
|\zeta(\tau)\rangle=e^{- k\tau/\hbar}e^{\zeta(-\tau)\mathscr{K}_{+}}e^{\eta \mathscr{Z}_3}e^{-\zeta(-\tau)^*\mathscr{K}_{-}}|k,0\rangle.
\end{equation}
Therefore, the time evolution of radial coherent states for the canonical  Dunkl-Klein-Gordon equation in curved spacetime, is expressed as
\begin{equation}
R_{nk}\left(x,\xi\right)=\left[\frac{2\Lambda^{k+\frac{1}{2}} \left(1 - |\xi|^2 \right)^{2k}}{ \Gamma\left(2k\right) } \right]^{1/2}\left( \sqrt{\Lambda}\, x \right)^{2k}e^{-ik}\frac{e^{\frac{i\Lambda{x^2}}{2}\left[\frac{\xi{e^{{-i\tau}/\hbar}}+1}{\xi{e^{{-i\tau}/\hbar}}-1}\right]}}{\left(1-\xi{e^{{-i\tau}/\hbar}}\right)^{2k}}.
\end{equation}
\begin{figure}[htbp]
    \centering
    \includegraphics[width=0.65\textwidth]{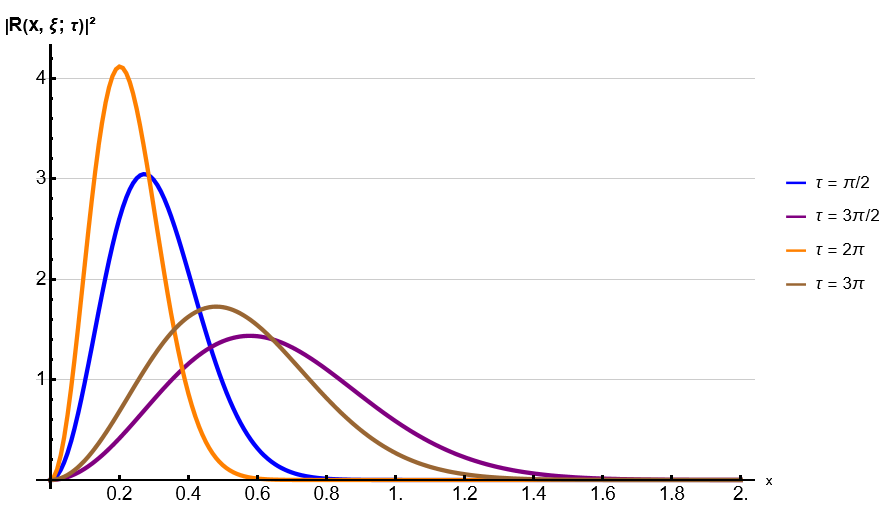}
    \caption{Time evolution of the normalized radial probability density $\left|R_{nk}(x, \xi, \tau)\right|^2$ for $n = 1$, $\alpha = \frac{1}{2}$, and $\xi = 0.5 + 0.2i$, evaluated at $\tau = \frac{\pi}{2}, \frac{3\pi}{2}, 2\pi, 3\pi$.}
\end{figure}
\begin{figure}[htbp]
    \centering
    \includegraphics[width=0.65\textwidth]{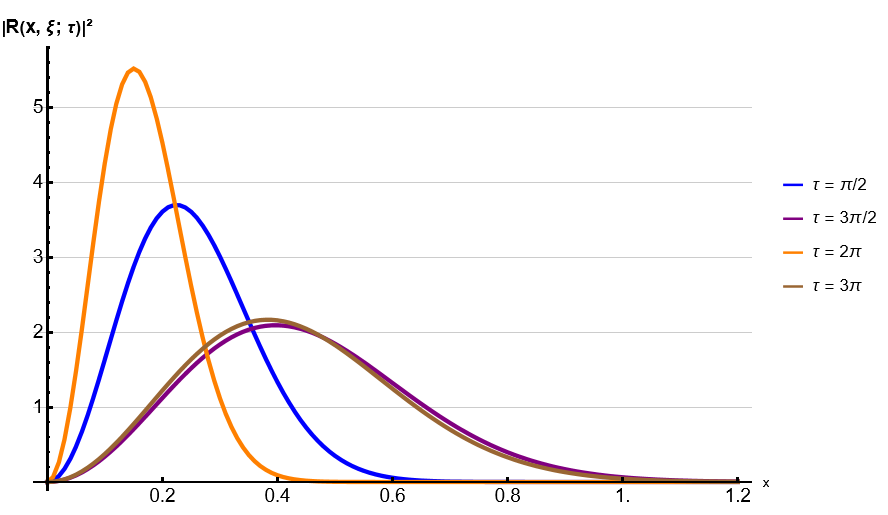}
    \caption{Time evolution of the normalized radial probability density $\left|R_{nk}(x, \xi, \tau)\right|^2$ for $n = 2$, $\alpha = \frac{1}{2}$, and $\xi = 0.5 + 0.2i$, evaluated at $\tau = \frac{\pi}{2}, \frac{3\pi}{2}, 2\pi, 3\pi$.}
\end{figure}
\begin{figure}[htbp]
    \centering
    \includegraphics[width=0.65\textwidth]{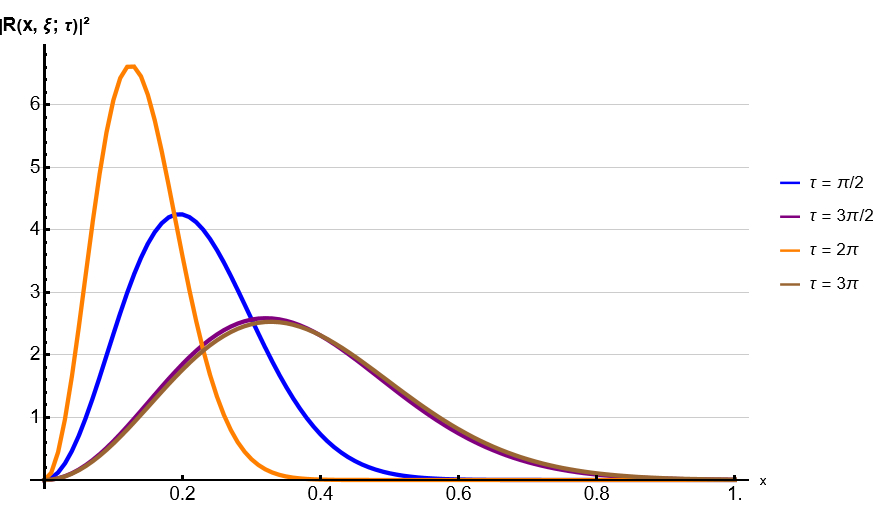}
    \caption{Time evolution of the normalized radial probability density $\left|R_{nk}(x, \xi, \tau)\right|^2$ for $n = 3$, $\alpha = \frac{1}{2}$, and $\xi = 0.5 + 0.2i$, evaluated at $\tau = \frac{\pi}{2}, \frac{3\pi}{2}, 2\pi, 3\pi$.}
\end{figure}
\begin{figure}[htbp]
    \centering
    \includegraphics[width=0.65\textwidth]{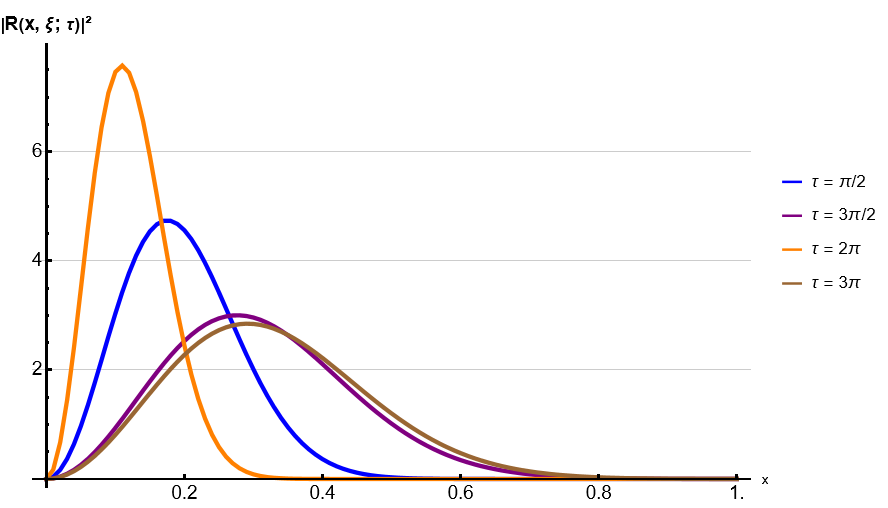}
    \caption{Time evolution of the normalized radial probability density $\left|R_{nk}(x, \xi, \tau)\right|^2$ for $n = 4$, $\alpha = \frac{1}{2}$, and $\xi = 0.5 + 0.2i$, evaluated at $\tau = \frac{\pi}{2}, \frac{3\pi}{2}, 2\pi, 3\pi$.}
\end{figure}
\FloatBarrier
\begin{figure}[H]
    \centering
    \includegraphics[width=0.65\textwidth]{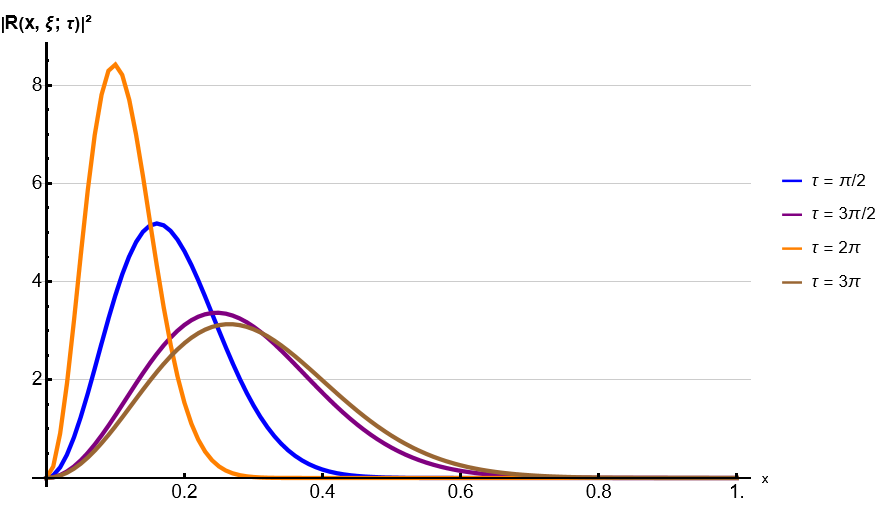}
    \caption{Time evolution of the normalized radial probability density $\left|R_{nk}(x, \xi, \tau)\right|^2$ for $n = 5$, $\alpha = \frac{1}{2}$, and $\xi = 0.5 + 0.2i$, evaluated at $\tau = \frac{\pi}{2}, \frac{3\pi}{2}, 2\pi, 3\pi$.}
\end{figure}

Figures corresponding to \( \alpha = \tfrac{1}{2} \) and \( \xi = 0.5 + 0.2i \) depict the time evolution of the normalized radial probability density \( |R_{nk}(x, \xi, \tau)|^2 \) for coherent states with quantum numbers \( n = 1, \ldots, 5 \), evaluated at the instants \( \tau = \tfrac{\pi}{2}, \tfrac{3\pi}{2}, 2\pi, 3\pi \). For each value of \( n \), the distributions exhibit a modulated Gaussian-like profile centered near the origin that evolves smoothly over time without significant delocalization. As \( \tau \) increases, the peak of the distribution shifts and its amplitude varies, reflecting the influence of the non-Hermitian evolution operator.

In addition, increasing the quantum number \( n \) results in narrower and more localized distributions toward smaller values of \( x \), with a pronounced increase in the peak amplitude. For instance, at \( \tau = 2\pi \), the maximum of the distribution for \( n = 1 \) is located around \( x \approx 0.3 \), with the function decaying to negligible values by \( x \approx 0.6 \), whereas for \( n = 5 \), the peak appears near \( x \approx 0.17 \), and the density rapidly vanishes beyond \( x \approx 0.4 \). This behavior indicates that higher-\( n \) states exhibit enhanced spatial localization and sharper modulation, characteristic of the quasi-stationary nature of the evolving coherent states. Similar behavior is observed for \( \alpha = \tfrac{3}{2} \) and \( \alpha = \tfrac{7}{2} \).

\section{Case: $a(x)=\frac{1-Rx^2}{1+Rx^2}$}
In this section, we examine the case \( a(x) = \frac{1 - R x^2}{1 + R x^2} \). Following an approach similar to that used previously, we determine the energy spectrum and the corresponding eigenfunctions. Thus, for very small values of the curvature constant \( R \) in \( a(x) \) (relative to \( E^2 - m^2 \)), Eq.~(\ref{EDSOE}) reduces to the form
\begin{equation}\label{EDDKSC}
\left[\frac{d^2}{dr^2}+\left( \frac{\alpha}{2} + \frac{3}{16} \right)\frac{1}{r^2}+\frac{1}{4}+\frac{E^2 - m^2+2R}{4\Theta{r}} \right]\Psi(r) = 0,
\end{equation}
where \( \Theta = \sqrt{4R\left(E^2 - m^2\right)} \). By applying the Schr\"odinger factorization to Eq.~(\ref{EDDKSC}), using the ansatz given in Eq.~(\ref{FACS}), we obtain the following complex coefficients
\begin{align}
\mathscr{A}_{SC}&=\pm\frac{1}{2}i , \hspace{0.5cm}\mathscr{C}_{SC}=\pm\frac{1}{2}i, \hspace{0.5cm} \mathscr{B}_{SC}=\pm\frac{E^2-m^2+2R}{4\Theta}i-1, \hspace{0.5cm} \mathscr{F}_{SC}=\pm\frac{E^2-m^2+2R}{4\Theta}i,\\
\mathscr{G}_{SC}&=\left(\frac{\alpha}{2} + \frac{3}{16}\right)-\left(i\frac{E^2-m^2}{4\Theta}-1\right)\left(i\frac{E^2-m^2}{4\Theta}\right).
\end{align}
Therefore, the differential equation (\ref{EDDKSC}) can be factorized as follows
\begin{equation}
\left[\mathscr{B}_{\pm}\pm1\right]\mathscr{B}_{\mp}=\mathscr{H}+\mathscr{R}\left[\mathscr{R}\pm i\right],
\end{equation}
where
\begin{equation}
\mathscr{H}=\left(\frac{\alpha}{2} + \frac{3}{16}\right), \hspace{0.5cm} \mathscr{R}=\left(\frac{E^2-m^2+2R}{4\Theta}\right),
\end{equation}
here, \( \mathscr{H} \) remains unchanged, while \( \mathscr{R} \) differs from \( \mathscr{L} \). Accordingly, in this case, the complex Schr\"odinger operators can be written as
\begin{equation}\label{OPSCAS}
\mathscr{B}_{\pm}= \mp r \frac{d}{dr} + \frac{1}{2}ir +i\left(\frac{E^2-m^2+2R}{4\Theta}\right).
\end{equation}
Eqs.~(\ref{EDDKSC}) and (\ref{OPSCAS}) allow us to express the following operators
\begin{equation}\label{OPKPM}
\mathscr{B}_{\pm}=\mp r \frac{d}{dr} + \frac{1}{2}ir +\mathcal{T}_3,
\end{equation}
where the operator $\mathcal{T}_3$, derived from Eq.~(\ref{EDDKSC}), is now defined as
\begin{equation}\label{OPT3SC}
\mathcal{T}_3=i\left[ r \frac{d^2}{d r^2} + \frac{\left(\frac{\alpha}{2} + \frac{3}{16}\right)}{r}  + \frac{1}{4} r \right]=-i\left(\frac{E^2-m^2+2R}{4\Theta}\right).
\end{equation}
As in the previous case, the operators \( \mathscr{B}_{\pm} \) and \( \mathcal{T}_3 \) also close an \( \mathfrak{su}(1,1) \) algebra; that is
\begin{equation}
[\mathcal{T}_3,\mathscr{B}_{\pm}]=\mp\mathscr{B}_{\pm}, \hspace{0.5cm} [\mathscr{B}_{-}, \mathscr{B}_{-}]=2\mathcal{T}_3.
\end{equation}
The energy spectrum corresponding to this choice of \( a(x) \) is obtained using the same procedure as in the previous section. The resulting group number \( k \) and the quadratic Casimir operator \( \mathbb{C}^2 \) coincide with those given in Eqs.~(\ref{CASC1}) and~(\ref{BARG}), respectively. Thus, from Eqs.~(\ref{BARG}) and~(\ref{OPT3SC}), we obtain the relation
\begin{equation}
\frac{1}{2}+\frac{\sqrt{1-8 \alpha}}{4}+n=-i\left(\frac{E^2-m^2+2R}{4\sqrt{4R(E^2 - m^2)}}\right),
\end{equation}
therefore, this final relation allows us to derive the energy spectrum given by
\begin{equation}
E^2 = m^2 - 2R \left[ \left( 2 + 2i\sqrt{2\alpha - \frac{1}{4}} + 4n \right)^2+1 \pm\sqrt{ \left[ \left( 2 + 2i\sqrt{2\alpha - \frac{1}{4}} + 4n \right)^2+1\right]^2-1}\right].
\end{equation}
from which we obtain the following table of values for the real and imaginary parts of \( E_{\pm} \)
\begin{table}[htbp]
\centering
\resizebox{\textwidth}{!}{%
\begin{tabular}{|c|c|c|c|c|c|c|c|}
\hline
\( \alpha \) & \( n \) & 0 & 1 & 2 & 3 & 4 & 5 \\
\hline
\multirow{2}{*}{\( \frac{1}{2} \)}
& \( E_+ \) & \( 3.297 - 4.223i \) & \( 3.432 - 12.114i \) & \( 3.452 - 20.072i \) & \( 3.458 - 28.053i \) & \( 3.460 - 36.041i \) & \( 3.461 - 44.034i \) \\
& \( E_- \) & \( 0.983 + 0.068i \) & \( 0.989 + 0.007i \)  & \( 0.995 + 0.002i \)  & \( 0.998 + 0.001i \)  & \( 0.999 + 0.0003i \) & \( 0.999 + 0.0002i \) \\
\hline
\multirow{2}{*}{\( \frac{3}{2} \)}
& \( E_+ \) & \( 6.468 - 4.107i \) & \( 6.582 - 12.096i \) & \( 6.611 - 20.067i \) & \( 6.621 - 28.051i \) & \( 6.626 - 36.040i \) & \( 6.628 - 44.033i \) \\
& \( E_- \) & \( 1.014 + 0.031i \) & \( 0.994 + 0.009i \)  & \( 0.996 + 0.003i \)  & \( 0.998 + 0.001i \)  & \( 0.999 + 0.001i \)  & \( 0.999 + 0.000i \) \\
\hline
\multirow{2}{*}{\( \frac{7}{2} \)}
& \( E_+ \) & \( 10.266 - 4.050i \) & \( 10.331 - 12.072i \) & \( 10.362 - 20.059i \) & \( 10.375 - 28.047i \) & \( 10.381 - 36.038i \) & \( 10.385 - 44.032i \) \\
& \( E_- \) & \( 1.012 + 0.011i \) & \( 0.999 + 0.008i \)  & \( 0.998 + 0.003i \)  & \( 0.998 + 0.001i \)  & \( 0.999 + 0.001i \)  & \( 0.999 + 0.000i \) \\
\hline
\end{tabular}
}
\caption{Real and complex values of \( E_+ \) and \( E_- \) for \( R = 1 \), \( m = 1 \), and \( \alpha = \frac{1}{2}, \frac{3}{2}, \frac{7}{2} \), with \( n = 0,\ldots,5 \).}
\end{table}
\FloatBarrier
\begin{figure}[H]
    \centering
    \includegraphics[width=0.70\textwidth]{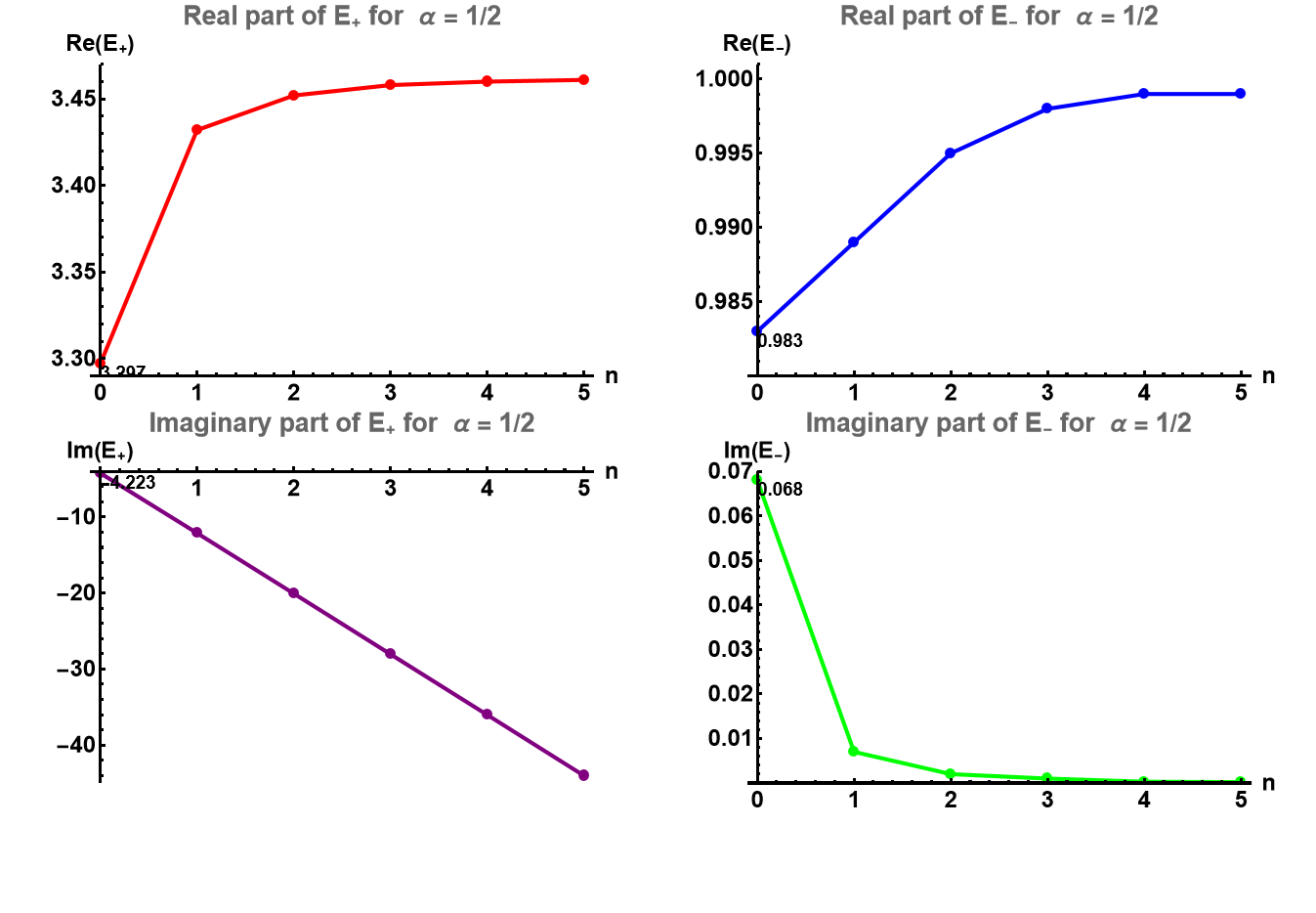}
    \caption{Energy spectrum \( E_+ \) and \( E_- \) for real and complex values for \( R = 1 \), \( m = 1 \), and \( \alpha = \frac{1}{2},\), with \( n = 0,\ldots,5 \).}
\end{figure}
Figure~9 shows the real and imaginary parts of the complex energy spectrum \( E_\pm \) as functions of the quantum number \( n \), for \( \alpha = \tfrac{1}{2} \). The upper plots display the real components of \( E_+ \) and \( E_- \), while the lower plots depict their corresponding imaginary parts: \( \operatorname{Im}(E_+) \) and \( \operatorname{Im}(E_-) \).

The \( E_+ \) branch exhibits an approximately constant real part in the range \( 3.3\text{--}3.5 \), along with a linearly decreasing imaginary part as \( n \) increases, indicating growing instability of the resonant modes. In contrast, the \( E_- \) branch remains close to unity in its real part and shows a rapidly vanishing imaginary component, signaling the presence of long-lived quasi-stationary states.

This spectral bifurcation between \( E_+ \) and \( E_- \) reveals the non-Hermitian nature of the deformed Dunkl--Klein--Gordon system and underscores the role of the underlying \( \mathfrak{su}(1,1) \) algebraic symmetry in structuring the spectrum. The complex nature of \( E_\pm \) precludes a conventional bound-state interpretation, highlighting the hybrid scattering--resonant character of the model.
\FloatBarrier
\begin{figure}[H]
    \centering
    \includegraphics[width=0.75\textwidth]{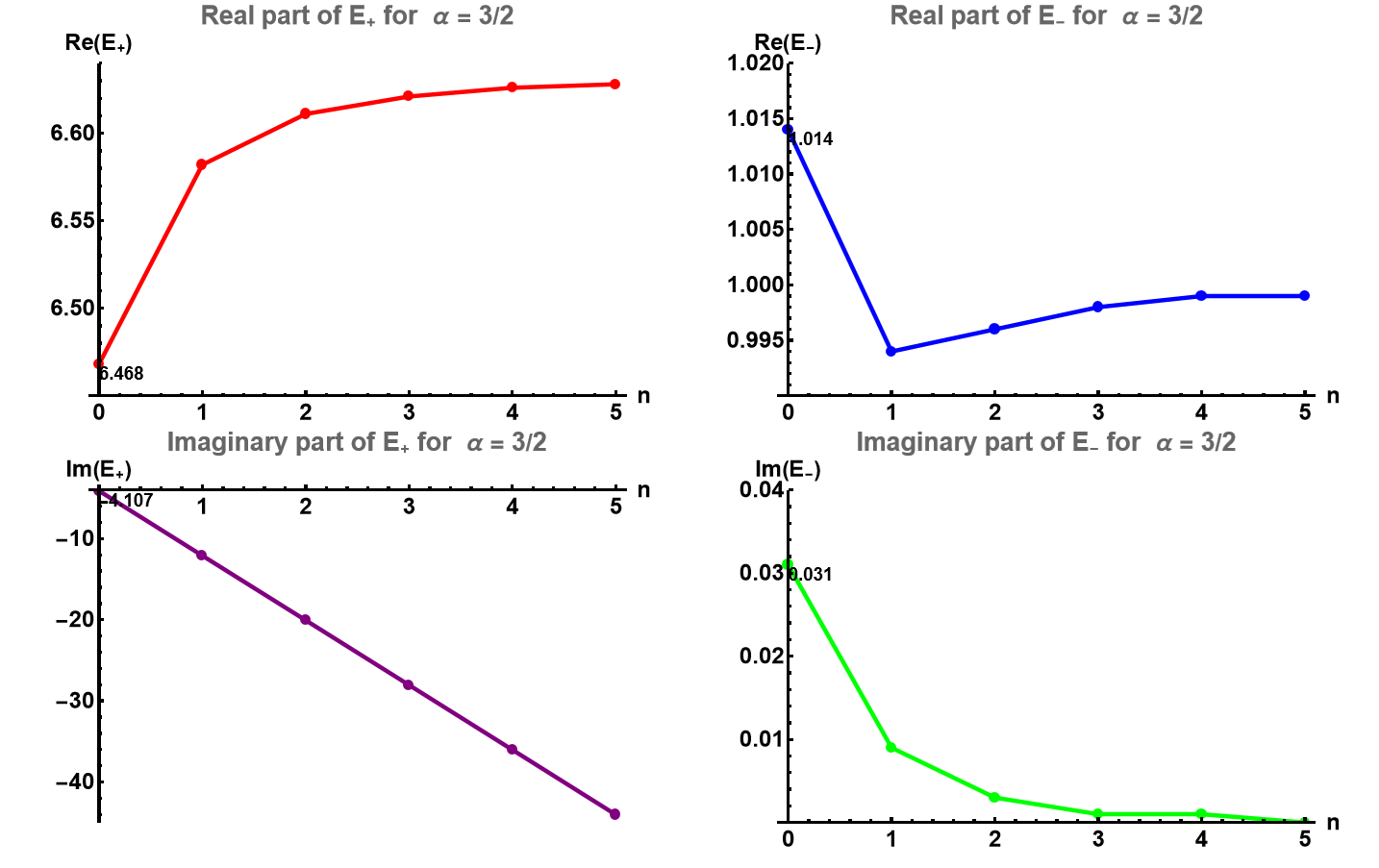}
    \caption{Energy spectrum \( E_+ \) and \( E_- \) for real and complex values for \( R = 1 \), \( m = 1 \), and \( \alpha = \frac{3}{2},\), with \( n = 0,\ldots,5 \).}
\end{figure}
Figure~10 shows the real and imaginary parts of the complex energy spectrum $E_\pm$ as functions of the quantum number $n$, for $\alpha = \frac{3}{2}$.The upper graphs show the real components of $E_+$ and $E_-$, while the lower graphs display their imaginary parts.The $E_+$ branch features nearly constant real values above 6.4 and a steadily decreasing imaginary part, indicating short-lived resonant modes. In contrast, $E_-$ exhibits a non-monotonic real part near unity and a rapidly decaying imaginary component, corresponding to quasi-stationary states. The spectral structure reflects the non-Hermitian nature of the system and the role of $\mathfrak{su}(1,1)$ symmetry in organizing the energy levels.
\FloatBarrier
\begin{figure}[H]
    \centering
    \includegraphics[width=0.75\textwidth]{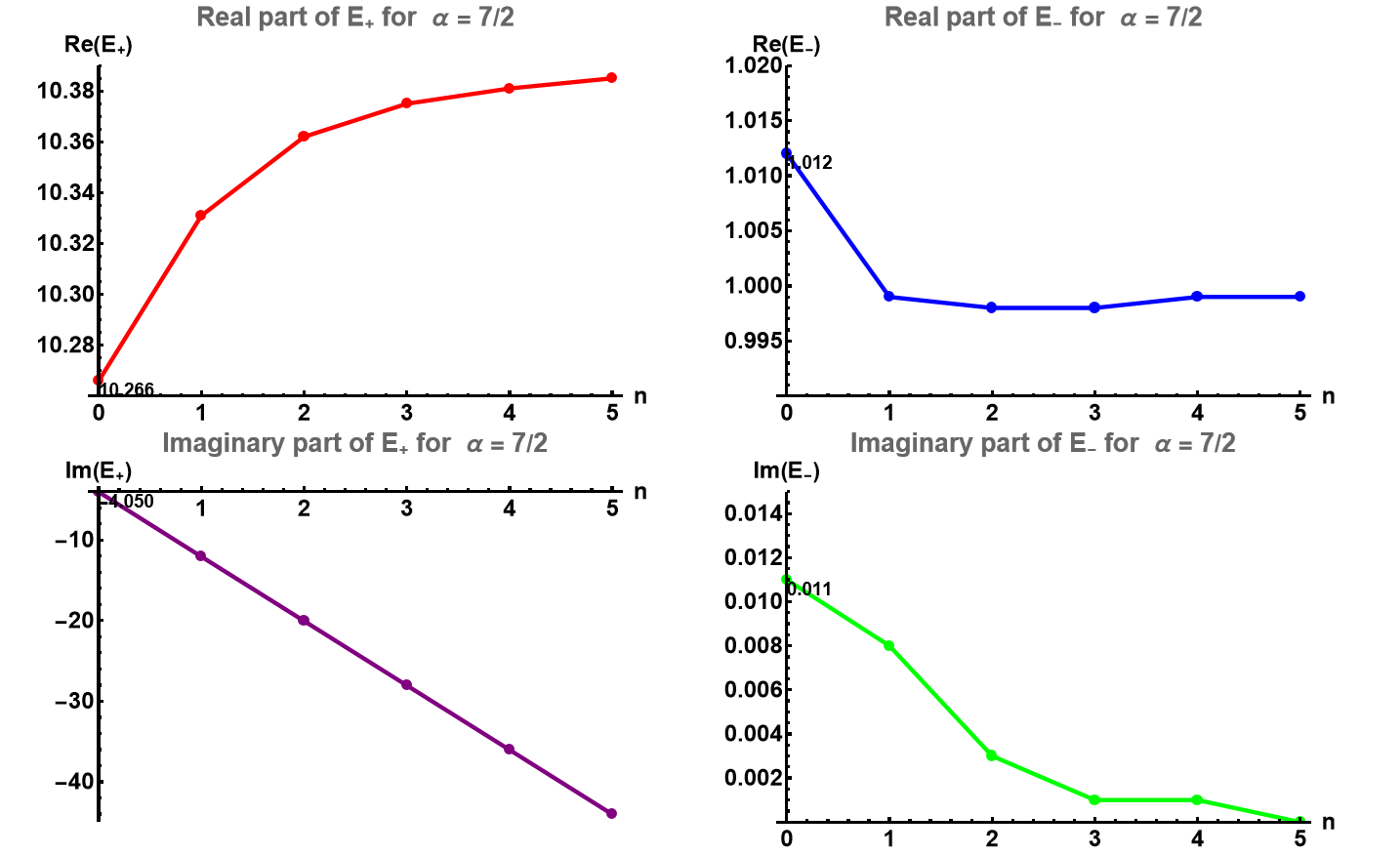}
    \caption{Energy spectrum \( E_+ \) and \( E_- \) for real and complex values for \( R = 1 \), \( m = 1 \), and \( \alpha = \frac{7}{2},\), with \( n = 0,\ldots,5 \).}
\end{figure}
Figure~11 shows the real and imaginary parts of the complex energy spectrum \( E_\pm \) as functions of the quantum number \( n \), for \( \alpha = \frac{7}{2} \). The upper graphs depict the real parts of \( E_+ \) and \( E_- \), while the lower graphs show their respective imaginary components. The \( E_+ \) branch features real parts around 10.3 and a linearly decreasing imaginary part, indicating increasingly unstable resonances. In contrast, the \( E_- \) branch remains close to unity with imaginary parts rapidly approaching zero, characterizing extremely long-lived modes. This strong spectral separation illustrates the influence of Dunkl deformation and the persistence of the underlying \( \mathfrak{su}(1,1) \) algebraic symmetry.

Finally, it can be concluded that as the Dunkl deformation parameter \( \alpha \) increases, the real part of \( E_+ \) shifts to higher values, while its imaginary part continues to decrease linearly with the quantum number \( n \), indicating progressively shorter-lived resonant states. In contrast, the real part of \( E_- \) remains close to unity across all values of \( \alpha \), and its imaginary component becomes increasingly small, reflecting enhanced stability of the associated quasi-bound states. This behavior results in a more pronounced spectral separation between the \( E_+ \) and \( E_- \) branches as \( \alpha \) grows. Consequently, the spectral bifurcation becomes sharper, underscoring the impact of the Dunkl deformation on the resonance structure.

\section{Case $a(x) = \frac{\sin(x\sqrt{R})}{x\sqrt{R}}$}
To complete our analysis, we consider the case $a(x) = \frac{\sin(x\sqrt{R})}{x\sqrt{R}}$ and apply the methodology developed in the preceding sections. Under this condition, Eq.~(\ref{EDSOE}) takes the form
\begin{equation}\label{EDTCS}
\left[\frac{d^2}{dr^2}+\left( \frac{\alpha}{2} + \frac{3}{16} \right)\frac{1}{r^2}+\frac{1}{4}+\frac{6\left(E^2 - m^2\right)+R}{24\Pi{r}} \right]G(r) = 0,
\end{equation}
here, \( \Pi = \sqrt{\frac{R\left(E^2 - m^2\right)}{3}} \). The complex coefficients obtained from the factorization of Eq.~(\ref{EDTCS}) using the ansatz given in Eq.~(\ref{FACS}) are
\begin{align}
\mathscr{A}_{TC}&=\pm\frac{1}{2}i , \hspace{0.5cm}\mathscr{C}_{TC}=\pm\frac{1}{2}i, \hspace{0.5cm} \mathscr{B}_{TC}=\pm\frac{6\left(E^2-m^2\right)+R}{24\Pi}i-1, \hspace{0.5cm} \mathscr{F}_{TC}=\pm\frac{6\left(E^2-m^2\right)+R}{24\Pi}i
\end{align}
\begin{equation}
\mathscr{G}_{TC}=\left(\frac{\alpha}{2} + \frac{3}{16}\right)-\left(i\pm\frac{6\left(E^2-m^2\right)+R}{24\Pi}-1\right)\left(i\pm\frac{6\left(E^2-m^2\right)+R}{24\Pi}\right).
\end{equation}
Thus, the differential equation~(\ref{EDTCS}) admits a factorized representation of the form
\begin{equation}
\left[\mathcal{X}_{\pm}\pm1\right]\mathcal{X}_{\mp}=\mathcal{Q}+\mathcal{I}\left[\mathcal{I}\pm i\right],
\end{equation}
where
\begin{equation}
\mathcal{Q}=\left(\frac{\alpha}{2} + \frac{3}{16}\right), \hspace{0.5cm} \mathcal{I}=\frac{6\left(E^2-m^2\right)+R}{24\Pi},
\end{equation}
for this particular choice of \( a(x) \), the complex Schr\"odinger operators are expressed as
\begin{equation}\label{OPSC3}
\mathcal{X}_{\pm}= \mp r \frac{d}{dr} + \frac{1}{2}ir +i\frac{6\left(E^2-m^2\right)+R}{24\Pi}.
\end{equation}
From Eqs.~(\ref{EDTCS}) and (\ref{OPSC3}), the following complex operators can be defined
\begin{equation}\label{OPKPM}
\mathcal{Z}_{\pm}=\mp r \frac{d}{dr} + \frac{1}{2}ir +\Gamma_3,
\end{equation}
the operator $\Gamma_3$ derived from Eq.~(\ref{EDTCS}) is now defined as
\begin{equation}\label{OPT3}
\Gamma_3=i\left[ r \frac{d^2}{d r^2} + \frac{\left(\frac{\alpha}{2} + \frac{3}{16}\right)}{r}  + \frac{1}{4} r \right]=-i\frac{6\left(E^2-m^2\right)+R}{24\Pi},
\end{equation}
where the operators \( \mathcal{Z}_{\mp} \) and \( \Gamma_3 \) also close an \( \mathfrak{su}(1,1) \) algebra. To determine the energy spectrum associated with this configuration, we note that the group number \( k \) and the quadratic Casimir operator \( \mathbb{C}^2 \) coincide with those previously identified in Eqs.~(\ref{CASC1}) and~(\ref{BARG}). Consequently, combining Eqs.~(\ref{OPTLA}) and~(\ref{OPT3}) yields the relation
\begin{equation}
\frac{1}{2}+\frac{\sqrt{1-8 \alpha}}{4}+n=-i\frac{6\left(E^2-m^2\right)+R}{24\Pi}.
\end{equation}
Therefore, the energy spectrum can be obtained directly from this final relation
\begin{equation}
E^2 = m^2 - \frac{R}{6} \left[ \left( 2 + 2i\sqrt{2\alpha - \frac{1}{4}} + 4n \right)^2+1 \pm\sqrt{ \left[ \left( 2 + 2i\sqrt{2\alpha - \frac{1}{4}} + 4n \right)^2+1\right]^2-1}\right].
\end{equation}
This leads to the following table containing the real and imaginary components of \( E_{\pm} \)
\begin{table}[htbp]
\centering
\resizebox{\textwidth}{!}{%
\begin{tabular}{|c|c|c|c|c|c|c|c|}
\hline
\( \alpha \) & \( n \) & 0 & 1 & 2 & 3 & 4 & 5 \\
\hline
\multirow{2}{*}{\( \frac{1}{2} \)}
& \( E_+ \) & \( 1.158 - 1.002i \) & \( 1.027 - 3.374i \) & \( 1.010 - 5.717i \) & \( 1.005 - 8.042i \) & \( 1.003 - 10.360i \) & \( 1.002 - 12.676i \) \\
& \( E_- \) & \( 0.998 + 0.006i \) & \( 0.999 + 0.001i \) & \( 1.000 + 0.0001i \) & \( 1.000 + 0.00005i \) & \( 1.000 + 0.00002i \) & \( 1.000 + 0.00001i \) \\
\hline
\multirow{2}{*}{\( \frac{3}{2} \)}
& \( E_+ \) & \( 1.001 + 0.003i \) & \( 1.957 - 3.390i \) & \( 1.932 - 5.721i \) & \( 1.924 - 8.044i \) & \( 1.921 - 10.361i \) & \( 1.919 - 12.676i \) \\
& \( E_- \) & \( 2.043 - 1.084i \) & \( 1.000 + 0.001i \) & \( 1.000 + 0.0002i \) & \( 1.000 + 0.00009i \) & \( 1.000 + 0.00004i \) & \( 1.000 + 0.00002i \) \\
\hline
\multirow{2}{*}{\( \frac{7}{2} \)}
& \( E_+ \) & \( 1.001 + 0.001i \) & \( 3.048 - 3.410i \) & \( 3.024 - 5.728i \) & \( 3.014 - 8.047i \) & \( 3.009 - 10.363i \) & \( 3.006 - 12.677i \) \\
& \( E_- \) & \( 3.096 - 1.119i \) & \( 1.000 + 0.001i \) & \( 1.000 + 0.0003i \) & \( 1.000 + 0.00012i \) & \( 1.000 + 0.00006i \) & \( 1.000 + 0.00004i \) \\
\hline
\end{tabular}
}
\caption{Real and complex values of \( E_+ \) and \( E_- \) for \( R = 1 \), \( m = 1 \), and \( \alpha = \frac{1}{2}, \frac{3}{2}, \frac{7}{2} \), with \( n = 0, \ldots, 5. \)}
\end{table}
\FloatBarrier
\begin{figure}[H]
    \centering
    \includegraphics[width=0.65\textwidth]{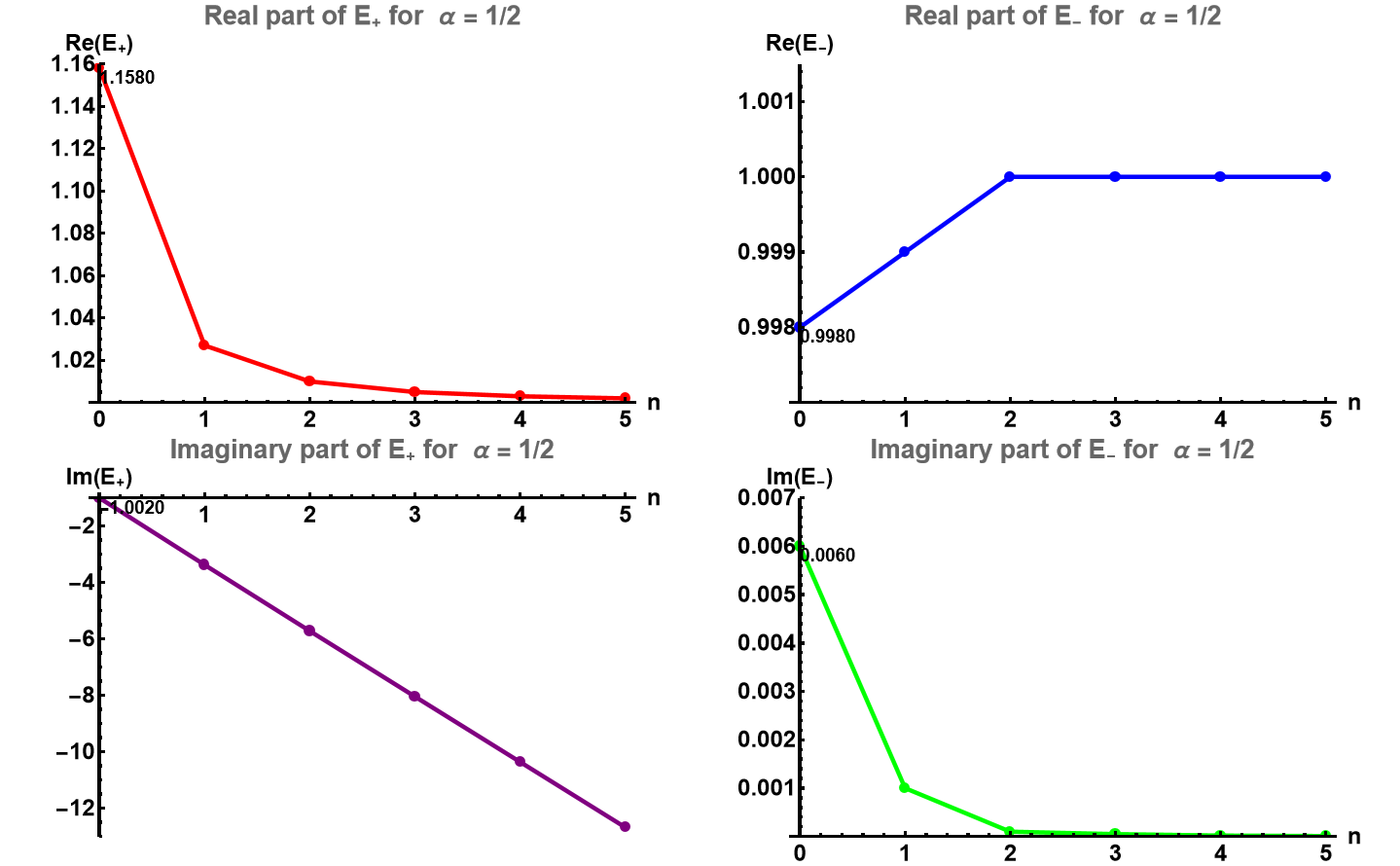}
    \caption{Energy spectrum \( E_+ \) and \( E_- \) for real and complex values for \( R = 1 \), \( m = 1 \), and \( \alpha = \frac{3}{2},\), with \( n = 0,\ldots,5 \).}
\end{figure}
Figure~12 illustrates the complex energy spectrum \( E_\pm \) as a function of the quantum number \( n \), computed from a deformed Klein--Gordon model with curvature-modified weight function \( a(x) = \frac{\sin(x\sqrt{R})}{x\sqrt{R}} \) and Dunkl parameter \( \alpha = \frac{1}{2} \). The upper graphs display the real parts of \( E_+ \) and \( E_- \), while the lower graphs show their respective imaginary components.

The \( E_+ \) branch features real parts that decrease sharply from approximately \( 1.16 \) to near \( 1 \), and imaginary parts that fall linearly with \( n \), revealing an increase in resonance decay rates. Meanwhile, the \( E_- \) branch remains tightly localized around unity in its real part and exhibits rapidly vanishing imaginary values, characterizing highly stable quasi-stationary modes.

This strong spectral asymmetry underscores the non-Hermitian structure of the Dunkl--Klein--Gordon system and the role of curvature and parity in shaping the energy landscape under \( \mathfrak{su}(1,1) \) symmetry constraints.
\FloatBarrier
\begin{figure}[H]
    \centering
    \includegraphics[width=0.65\textwidth]{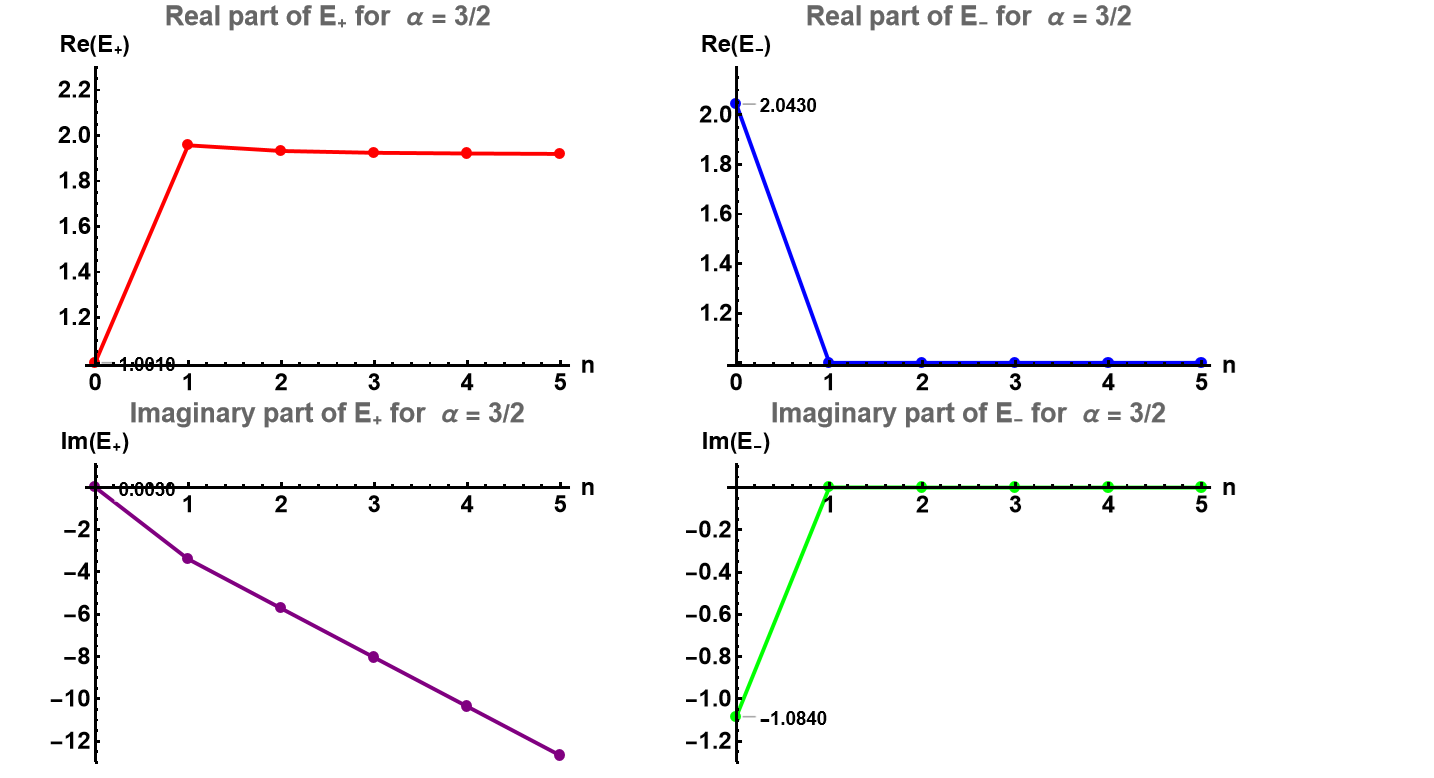}
    \caption{Energy spectrum \( E_+ \) and \( E_- \) for real and complex values for \( R = 1 \), \( m = 1 \), and \( \alpha = \frac{3}{2},\), with \( n = 0,\ldots,5 \).}
\end{figure}
Figure 13 shows the evolution of the real and complex energy spectrum \( E_\pm \) as a function of the quantum number \( n \), for \( \alpha = \frac{3}{2} \), derived from the expression involving the curvature-adjusted weight function \( a(x) = \frac{\sin(x\sqrt{R})}{x\sqrt{R}} \).

The upper graphs show the real parts of \( E_+ \) and \( E_- \), while the lower graphs represent their respective imaginary components. Compared to the case \( \alpha = \frac{1}{2} \), a spectral inversion occurs: the \( E_+ \) branch starts near unity and quickly stabilizes around 1.9, while its imaginary part becomes increasingly negative as \( n \) increases, indicating greater resonance decay.

In contrast, the \( E_- \) branch starts with a real part greater than 2 and a large negative imaginary component, but quickly converges to a nearly real value that remains essentially constant, indicating the onset of long-lasting quasi-stationary behavior. This inversion reflects a transition in the spectral structure driven by the interaction between Dunkl's deformation and curvature effects.
\FloatBarrier
\begin{figure}[H]
    \centering
    \includegraphics[width=0.65\textwidth]{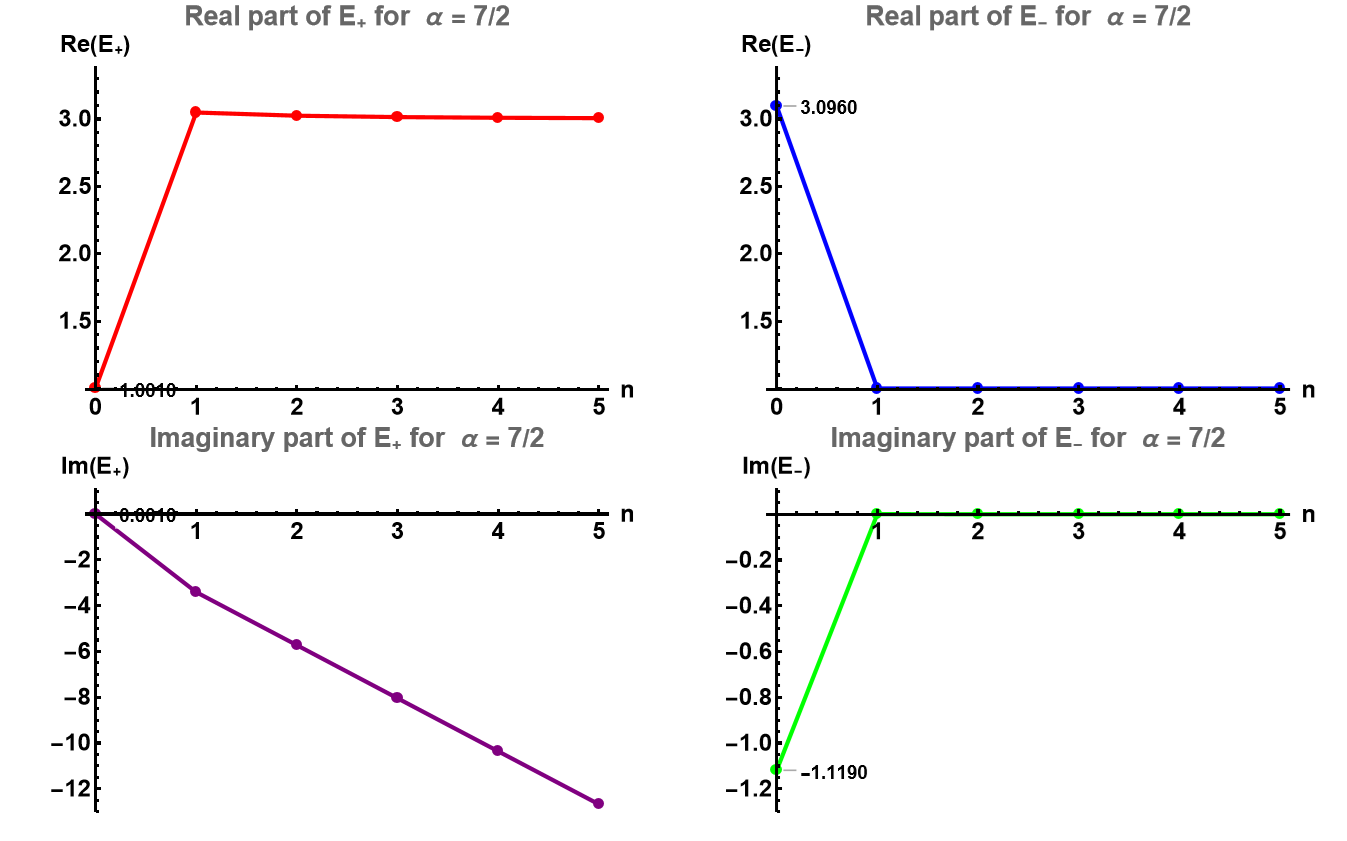}
    \caption{Energy spectrum \( E_+ \) and \( E_- \) for real and complex values for \( R = 1 \), \( m = 1 \), and \( \alpha = \frac{3}{2},\), with \( n = 0,\ldots,5 \).}
\end{figure}
Figure~14 illustrates the behavior of the real and imaginary parts of the complex energy spectrum \( E_\pm \) as a function of the quantum number \( n \), for the case \( \alpha = \frac{7}{2} \), using the curvature-modified weight function \( a(x) = \frac{\sin(x\sqrt{R})}{x\sqrt{R}} \). The upper plots display the real parts of \( E_+ \) and \( E_- \), while the lower graphs show their respective imaginary components.

For the \( E_+ \) branch, the real part starts near unity and quickly stabilizes around 3.0 as \( n \) increases, whereas its imaginary part decreases almost linearly, indicating an enhanced decay rate of the associated resonant modes. In contrast, the \( E_- \) branch begins at \( n = 0 \) with a large real part (\( \approx 3.096 \)) and a significantly negative imaginary component (\( \approx -1.119 \)), but both collapse abruptly for \( n \geq 1 \) toward nearly constant values: the real part stabilizes around unity and the imaginary part becomes negligible.

This sudden transition suggests the emergence of long-lived quasi-stationary states. Altogether, these trends highlight how large values of the parameter \( \alpha \) drive a clear separation between rapidly decaying resonances and stable modes, a hallmark of the non-Hermitian \( \mathfrak{su}(1,1) \) symmetric structure of the system.

In summary, increasing the Dunkl deformation parameter \( \alpha \) induces a notable reorganization of the complex energy spectrum. The real part of the \( E_+ \) branch rises steadily with \( \alpha \), while its imaginary component becomes increasingly negative as \( n \) increases, signaling faster decay rates of the resonant modes. Conversely, the \( E_- \) energies remain tightly clustered around unity in their real parts, with imaginary components that rapidly approach zero, characteristic of highly stable quasi-bound states. As a result, the gap between \( E_+ \) and \( E_- \) becomes more pronounced, evidencing the deepening spectral asymmetry introduced by the Dunkl deformation and its effect on the resonance--continuum structure of the model.

Analogous results hold for the coherent states and their time evolution in the cases \( a(x) = \frac{1 - R x^2}{1 + R x^2} \) and \( a(x) = \frac{\sin(x\sqrt{R})}{x\sqrt{R}} \). While the effective radial equations for all three configurations reduce to a common differential form solvable via generalized Laguerre polynomials, this mathematical similarity does not translate to physical equivalence.

The critical difference emerges in the energy spectrum \( E_n^2 \), which depends explicitly on the functional form of \( a(x) \), modifying both the Bargmann index \( k \) and the algebraic generators governing the dynamics. Consequently, despite the radial wave functions analogous analytic structure, the coherent states display markedly different localization patterns and time evolution. These variations highlight the distinct roles of geometry, curvature, and \(\mathfrak{su}(1,1)\) symmetry in each realization of the Dunkl-Klein--Gordon system.

\section{Concluding Remarks}
Using representation-theory techniques for the $\mathfrak{su}(1,1)$ symmetry algebra, we derive the complex eigenfunctions of the Dunkl-Klein-Gordon problem in closed form, expressed through Laguerre polynomials. Within this formalism, we reproduce the complex energy spectrum reported in Ref.~\cite{Seda3} for $a(x) = e^{-R x^{2}}$. We further obtain the spectra for two other profiles, $a(x) = \tfrac{1 - R x^{2}}{1 + R x^{2}}$ and $a(x) = \tfrac{\sin(x\sqrt{R})}{x\sqrt{R}}$.Although the resulting energy spectrum is complex-indicating that the system is non-Hermitian-the algebraic \( \mathfrak{su}(1,1) \) symmetry remains valid. This symmetry does not require the Hamiltonian to be Hermitian, as it is defined purely through the commutation relations between operators that satisfy the Lie algebra structure. In the present context, involving curvature-induced potentials and Dunkl-type deformations, the \( \mathfrak{su}(1,1) \) algebra serves as a powerful analytical framework for deriving exact solutions, even when conventional notions of energy conservation do not strictly apply. Finally, we construct the radial $SU(1,1)$ coherent states and study their time evolution, providing graphical representations of both  cases.

Our analysis was restricted to the even-parity sector and to the regime in which the curvature constant $R$ was much smaller than the kinetic energy of the system. It's also worth mentioning that the study equation arose from a matrix-operator algebra constructed with Dirac gamma matrices and a universal length scale as a measure of spatial curvature based on the Dunkl operator, thereby avoiding the introduction of spin connections in the Dirac equation.

\end{document}